\documentclass[10pt]{iopart}
\usepackage{epsfig}
\def\Journal#1#2#3#4{{#1} {\bf #2}, #3 (#4)}

\def\ZPC{{\em Z. Phys.} C}

\begin{document}
\noindent
Talk presented at the International Workshop
``On the Physics of the Quark-Gluon Plasma``,
Palaiseau, France, Sept. 2001.

\title{Strange Particle Production from SIS to LHC}
\author{H. Oeschler$^1$,
J. Cleymans$^2$, 
and K. Redlich$^{3,4,5}$
}
\address{
$^1$Institut f\"ur Kernphysik, Darmstadt University of Technology, 
D-64289 Darmstadt, Germany\\
$^2$Department  of  Physics,  University of Cape Town,
Rondebosch 7701, South Africa\\
$^3$Gesellschaft f\"ur Schwerionenforschung, D-64291 Darmstadt, Germany\\
$^4$Institute for Theoretical Physics, University of Wroc\l aw,
PL-50204  Wroc\l aw, Poland\\
$^4$now: Theory Division, CERN, CH-1211 Geneva 23, Switzerland\\}
\begin{abstract}
A review of meson emission in heavy ion collisions at incident energies
from SIS up to collider energies is presented.
A statistical model assuming chemical equilibrium and local strangeness 
conservation (i.e.~strangeness conservation per collision)
explains most of the observed features.

Emphasis is put onto the study of $K^+$ and $K^-$ emission at low incident
energies. 
In the framework of this statistical model it is shown that the experimentally
observed equality of $K^+$ and $K^-$ rates at ``threshold-corrected'' energies
$\sqrt{s} - \sqrt{s_{th}}$ is due to a crossing of two excitation functions.
Furthermore, the independence of the $K^+$ to $K^-$ ratio on the number of
participating nucleons observed between SIS  and RHIC is consistent with
this model.

It is demonstrated that the $K^-$ production at SIS energies occurs
predominantly via strangeness exchange and this channel
is approaching chemical equilibrium.
The observed maximum in the $K^+/\pi^+$ excitation function 
is also seen in the ratio of strange to non-strange particle production.
The appearance of this maximum around 30 $A\cdot$GeV
is due to the energy dependence of the chemical freeze-out parameters
$T$ and $\mu_B$.  
\end{abstract}

\section{Introduction}
Central heavy ion collisions
at relativistic incident energies represent an ideal tool to study
nuclear matter at high temperatures.
Particle production is -- at all incident energies -- a key quantity to extract
information on the properties of nuclear matter under these extreme conditions.
Particles carrying strangeness have turned out to be very valuable messengers.

A specific purpose of this paper is the presentation of the evolution
of strange particle production over a large range of incident energies.
The data  at low incident energies, i.e.~at and below the production 
threshold in $NN$ collisions will be shown in more detail as these results
are less presented in other works. 
Many results are shown together with a theoretical interpretation.
The attempts to describe particle production yields with 
statistical models~\cite{Cley,PBM,CLEY98,CLE99,CLE00,becattini,pbm99,PBM_RHIC}
have turned out to be very successful over this large domain of incident 
energies.

\section{General Trends}

\subsection{Production of $K^+$ and $K^-$ from SIS to RHIC}

At incident energies around 1 $A\cdot$GeV pion and kaon production is very 
different: Pions can be produced by direct $NN$ collisions but kaons not.
The threshold for $K^+$ production in $NN$ collisions is 1.58 $A\cdot$GeV
and only collective effects can accumulate the energy needed to produce
a $K^+$ together with a $\Lambda$ (or another strange particle) 
for strangeness conservation.

\begin{figure} 
\begin{minipage}[t]{5.3cm}
\mbox{\epsfig{width=9.0cm,file=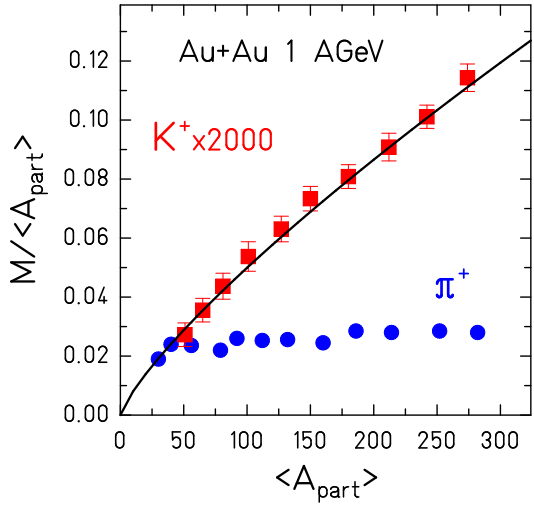}}
\end{minipage}
\hfill
\begin{minipage}[b]{7.6cm}
\vspace*{-3.6cm}
\caption{The multiplicity of $K^+/A_{part}$ rises strongly
with $A_{part}$ in contrast to the pion multiplicity~\protect\cite{Mang}.
This rise can be described by the statistical model
including local strangeness conservation (see text).}
\label{Apart}
\end{minipage}
\end{figure}

The measured multiplicities for pions and $K^+$ differ 
evidently by orders of magnitude. They exhibit a further 
very pronounced contrast: While the pion multiplicity per number of
participating nucleons $A_{part}$ remains constant 
with $A_{part}$, the $K^+$ 
multiplicity per $A_{part}$ rises strongly (Fig.~\ref{Apart}).
The latter observation seems to be in conflict with a thermal
interpretation, which -- in a naive view -- should give multiplicities per
mass number $A$ being constant.

Usually, the particle number densities or
the multiplicities per $A_{part}$, here for pions, are described
in a simplified way by a Boltzmann factor
$$\frac{M_{\pi}}{A_{part}}\sim \exp \left(-\frac{<E_{\pi}>}{T}\right),
$$with the temperature $T$ and the total energy $<E_{\pi}>$.

The production of strange particles has to fulfil strangeness
conservation. The attempt to describe the measured particle ratios
including strange hadrons at AGS and SPS using a strangeness chemical potential
$\mu_S$ is quite successful~\cite{Cley,PBM,pbm99,PBM_RHIC}.
However, this grand-canonical treatement is not sufficient,
if the number of produced strange particles is small.
Then, a statistical model has to take care of {\it exact
strangeness conservation} in each reaction as introduced in
~\cite{Hagedorn}.
This is done by taking into account
that e.g.~together with each $K^+$ a $\Lambda$ or another strange particle
is produced:
$$\frac{M_{K^+}}{A_{part}}\sim \exp \left(-\frac{<E_{K^+}>}{T}\right)
\left[g_{\Lambda}V \int {d^3p\over (2\pi)^3}
\exp\left(-{{(E_{\Lambda}-\mu_B)}\over T}\right)\right],
$$where $T$ is the temperature, $\mu_B$ the baryo-chemical potential,
$g_i$ the degeneracy factors, $V$ the production volume for making the
associate pair (see~\cite{CLE99,CLE00}) and $E_i$ the total energies.
We note that this volume is not identical to the volume of the system at freeze
out.
The volume parameter $V$ is
taken as $r_0^3 A_{part}$ with a common $r_0$ for all systems and all
incident energies.

This formula, simplified for demonstration purposes,
neglects other combinations leading to the production
of $K^+$ as well as the use of Bose-Fermi distributions, which are all
included in the computation.
The corresponding formula for $K^-$ production 
$$\frac{M_{K^-}}{A_{part}}\sim \exp \left(-\frac{<E_{K^-}>}{T}\right)
\left[g_{K^+}V \int {d^3p\over (2\pi)^3}
\exp\left(-{E_{K^+}\over T}\right)\right].
$$is similar, but does not
depend on $\mu_B$. This point will become important later on.

These formulae lead to
a reduction of $K^+$ and $K^-$ yields as compared to the numbers calculated
without exact strangeness conservation~\cite{CLE99,CLE00}.
Two extreme conditions can be seen from these equations.
In the limit of a small number of strange particles, 
the additional term (due to the parameter $V$)
leads to a linear rise of $M_{K^+}/A_{part}$
while $M_{\pi}/A_{part}$ remains constant.
This is in remarkable agreement with the experimental
observations shown in Fig.~\ref{Apart}. 
For very high temperatures or very large volumina, the terms in brackets
approach unity (see Ref.~\cite{CLE99}) and the formulae coincide
with the grand-canonical procedure. This is much better seen in the exact
formulae using modified Bessel functions~\cite{CLE99,CLE00,Hamieh}.

\begin{figure}[h]
\begin{minipage}[t]{5.1cm}
\mbox{\epsfig{width=9.4cm,file=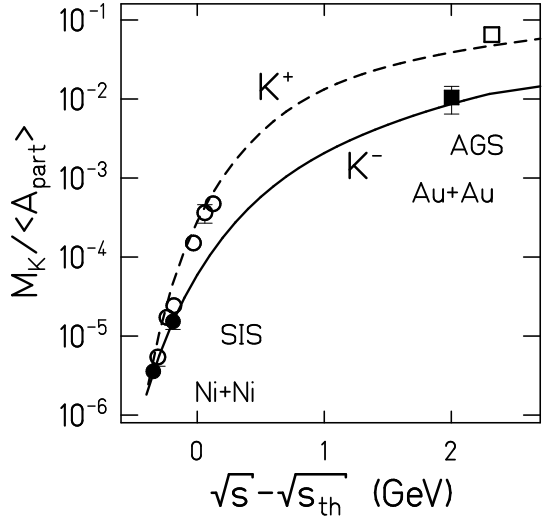}}
\end{minipage}
\hfill
\begin{minipage}[b]{7.8cm}
\vspace*{-3.8cm}
\caption{Calculated $K^+/A_{part}$ and $K^-/A_{part}$ ratios
in the statistical model as a function of $\sqrt{s} - \sqrt{s_{th}}$
for Ni+Ni collisions.
The points are results for Ni+Ni collisions at SIS energies
\protect\cite{Barth,Marc}
and Au+Au at 10.2 $A\cdot$GeV (AGS)~\protect\cite{Ahle}. 
At AGS energies the influence of the system mass is negligible.}
\label{KP_KM_sthr_therm}
\end{minipage}
\end{figure}

At low incident energies, the particle ratios (except $\eta/\pi_0$)
are well described using the canonical approach~\cite{CLE99}.
Surprisingly, even the measured $K^+/K^-$ ratio is described and 
this ratio does not depend on the choice of the volume term $V$.
It should be noted that the statistical model uses normal masses of the
particles while many transport calculation \cite{Cassing}
have to reduce the $K^-$ mass (as expected for kaon in the nuclear medium)
in order to describe the measured yields. 
It is therefore of interest to see how the results of the statistical model
appear in a representation where the $K^+$ and $K^-$ multiplicities are given
as a function of $\sqrt{s} - \sqrt{s_{th}}$
as shown in Fig.~\ref{KP_KM_sthr_therm}.
The choice of the x-axis follows the idea to correct for the different energy
thresholds to produce $K^+$ in $NN$ collisions ($\sqrt{s_{th}}$ = 2.548 GeV)
and $K^-$ ($\sqrt{s_{th}}$ = 2.87 GeV). 
It turned out that in this representation the measured yields of $K^+$ and 
$K^-$ close to threshold in heavy ion collisions are about equal while 
they differ in pp collisions by factors of 10 -- 100 \cite{Laue}.

Figure~\ref{KP_KM_sthr_therm} demonstrates that at
values of $\sqrt{s} - \sqrt{s_{th}}$ less than zero,
the excitation functions for $K^+$ and $K^-$ cross leading to 
the observed equality of 
$K^+$ and $K^-$ at SIS energies.
The yields differ at AGS energies by a factor of five.
The difference in the rise of the two excitation functions
can be understood by the formulae given above.
The one for $K^+$ production contains ($E_{\Lambda}-\mu_B$) while the other
has $E_{K^+}$ in the exponent of the second term. 
As these two values are different, the
excitation functions, i.e.~the variation with $T$, exhibit a different rise.

\begin{figure}
\begin{minipage}[t]{5.3cm}
\mbox{\epsfig{width=9.5cm,file=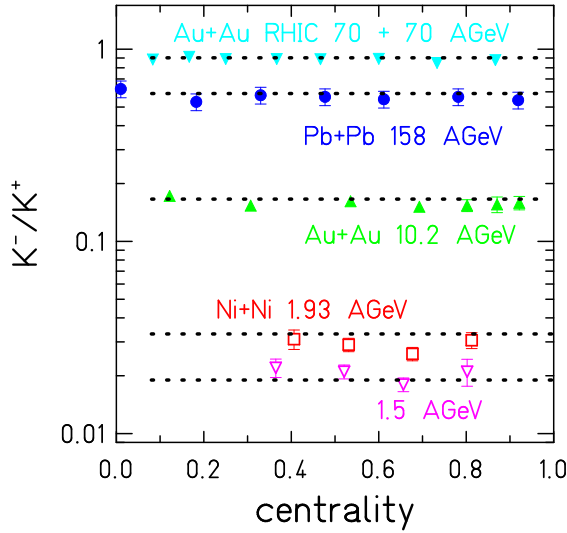}}
\end{minipage}
\hfill
\begin{minipage}[b]{7.8cm}
\vspace*{-5.8cm}
\caption{The $K^+/K^-$ ratio appears to be constant of a function of centrality
from SIS up to RHIC energies. The dotted lines represent the predictions of 
the statistical model. Data from \protect\cite{Marc,Ahle,Star}.}
\label{KPKM_SIS_RHIC}
\end{minipage}
\end{figure}

Furthermore, the two formulae predict that the $K^+/K^-$ ratio for a given 
collision should not vary with centrality as $V$ cancels in the ratio.
Indeed, this has been observed in Au+Au/Pb+Pb collisions between 1.5 $A\cdot$GeV
and RHIC energies~\cite{Marc,Ahle,AF,Dunlop,Star}
as shown in Fig.~\ref{KPKM_SIS_RHIC}.
This independence of centrality is most astonishing
as one expects at low incident energies an influence
of the different thresholds and the density variation with centrality.
At 1.93 $A\cdot$GeV the $K^+$ production is above and
the $K^-$ production below their respective $NN$ thresholds.
             
Transport-model calculations show clearly that
strangeness equilibration requires a time interval of
40 -- 80 fm/$c$~\cite{Koch86,Brat00}.
On the other hand, the statistical models assuming chemical equilibration
are quite successful in describing the particle yields, 
including strange particles.

In case of the $K^+$ production,
no strong absorptive channel seems to be available which could lead to 
chemical equlibration.  
For $K^-$ production the situation is quite different.
At low incident energies strange quarks are found only in
a few hadrons. The $\bar s$ quark is essentially only in $K^+$, while
the $s$ quark will be shared between $K^-$ and $\Lambda$ (or other hyperons).
This sharing of the $s$ quark might be in chemical equlibrium as
the reactions $$\pi^0 + \Lambda \rightleftharpoons p + K^- \quad \rm{or} \quad
\pi^- + \Lambda \rightleftharpoons n + K^-$$
 are strong and have only
slightly negative Q-values of -176 MeV.

The idea that the $K^-$ yield is dominated by strangeness exchange via
the $\pi^- + \Lambda$ channel has been suggested by~\cite{Ko84}
and has been demonstrated quantitavely in a recent theoretical
study~\cite{Hart01}. The direct $K^+ K^-$ pair production via baryon-baryon
collisions has negligible influence as these $K^-$ are absorbed entirely.
In these transport-model calculations the strangeness exchange is approaching
equilibrium but does not fully reach it~\cite{Hart01}.

If these reactions are the dominating channel, the law of mass action
might be applied giving for the respective concentrations~\cite{oeschler_s2000}
$$\frac{[\pi] \cdot [\Lambda]}{[K^-] \cdot N} \, = \, \kappa .$$
As the number of $K^-$ relative to $\Lambda$ is small, $[\Lambda]$ can be
approximated by $[K^+]$ and rewriting gives
$$\frac{[K^-]}{[K^+]} \propto M(\pi^0 + \pi^-)/A_{part}.$$

\begin{figure}[t] 
\mbox{\epsfig{width=4.2cm,file=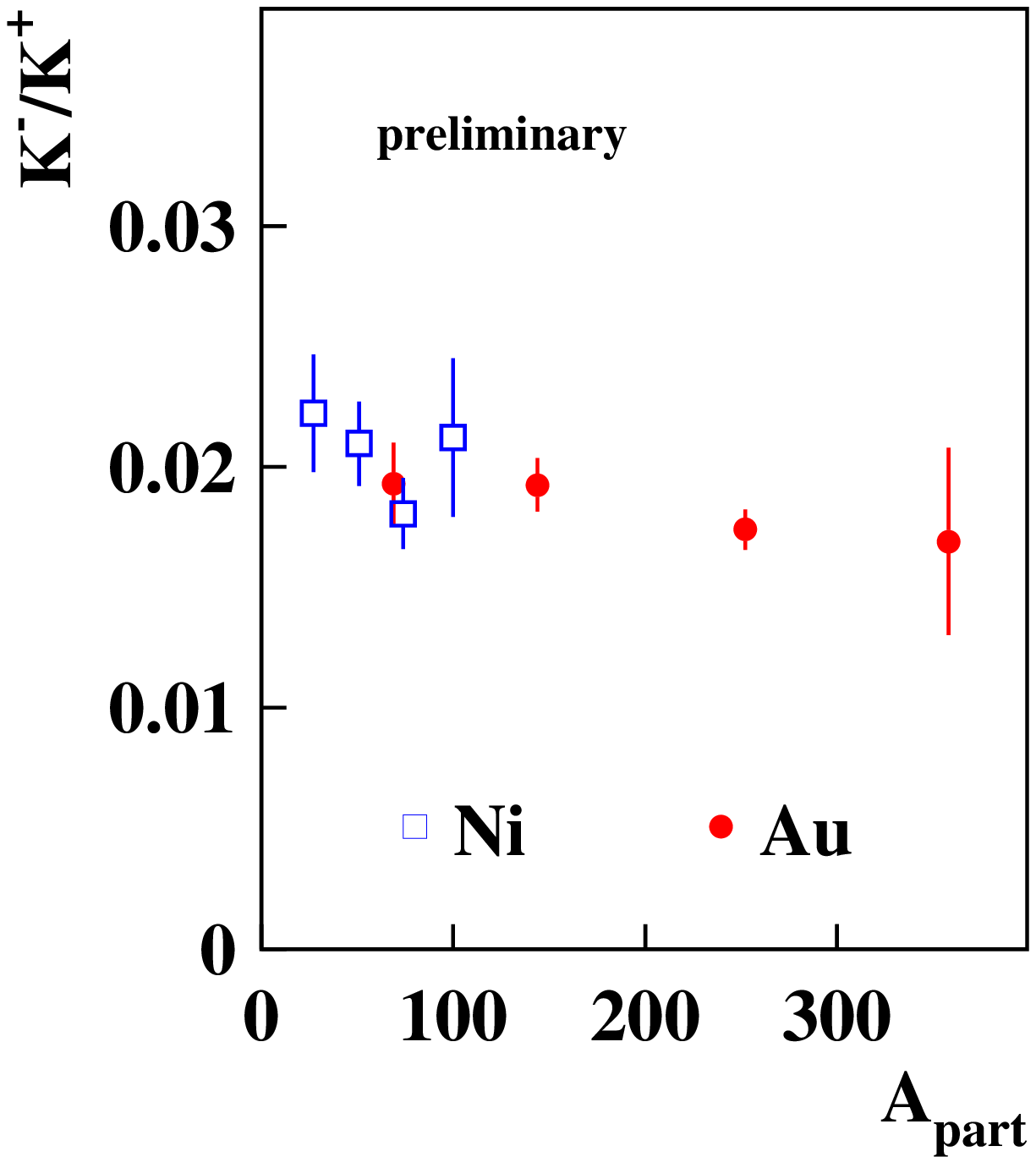}}
\mbox{\epsfig{width=4.2cm,file=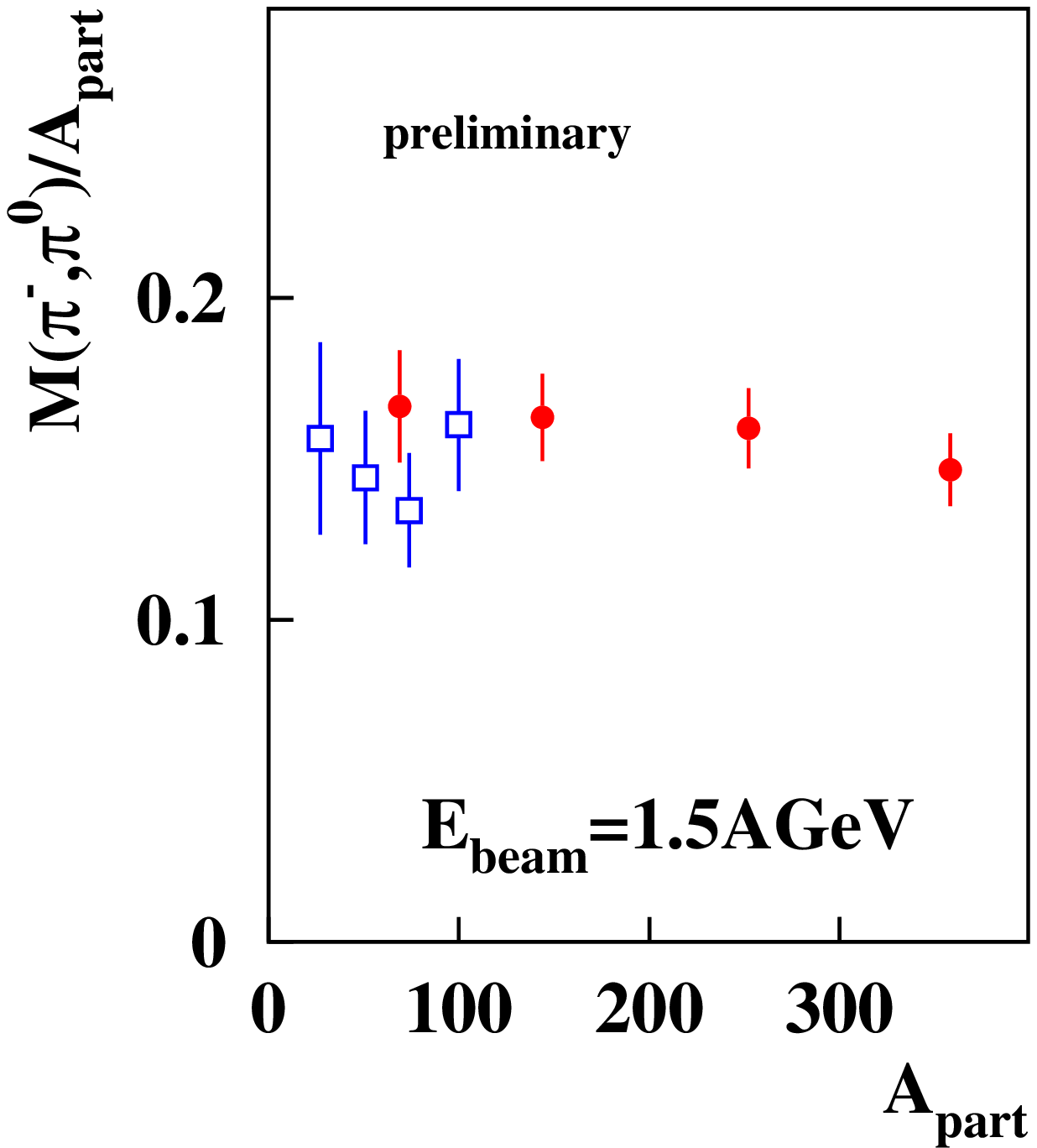}}
\mbox{\epsfig{width=4.2cm,file=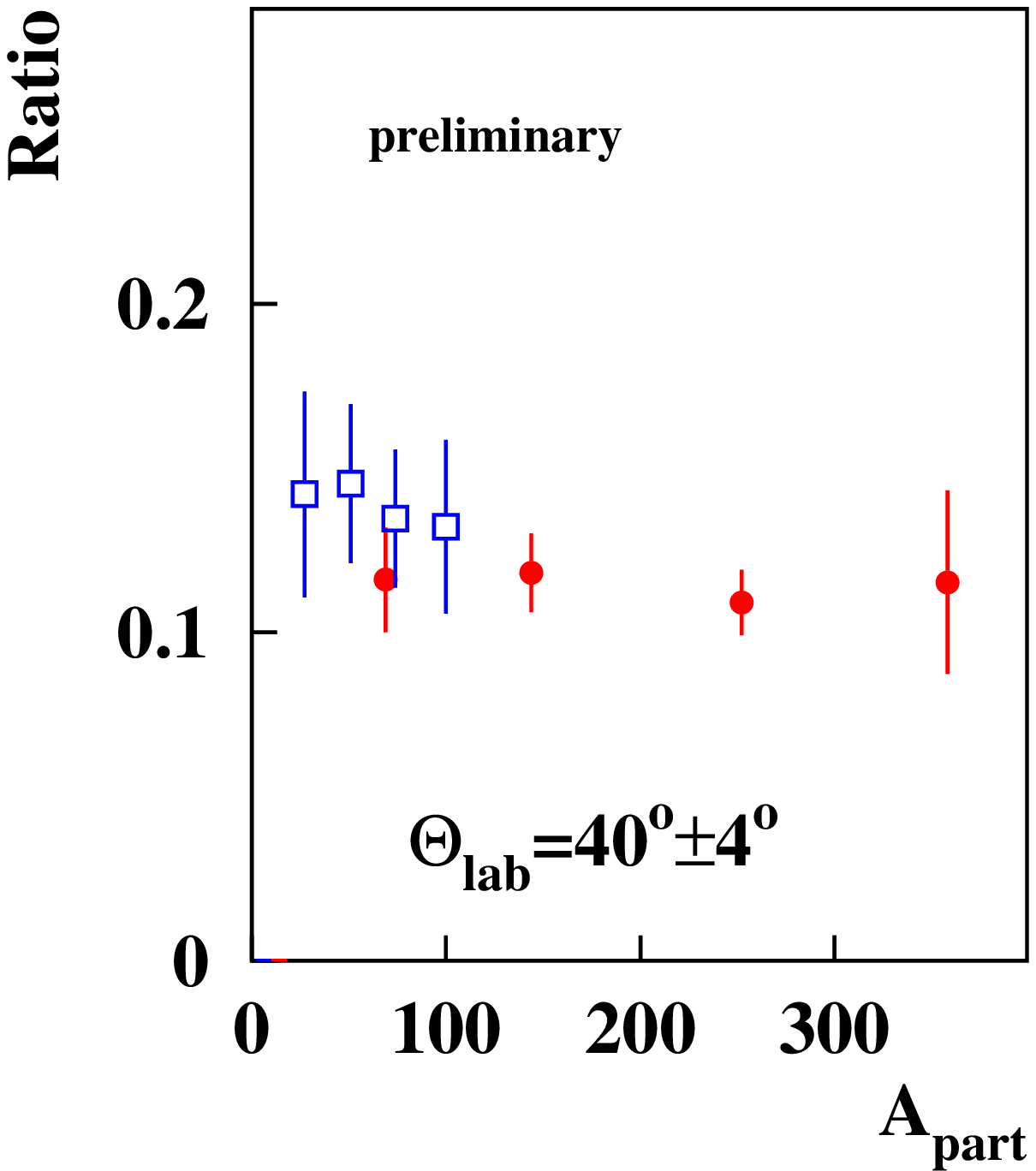}}
\caption{Measured~$K^-/K^+$ ratio, $M(\pi^0 + \pi^-)/A_{part}$ and~the double
ratio $([K^-]/[K^+])/([M(\pi^0 + \pi^-)]/A_{part})$  as a function of 
$A_{part})$ both for Ni+Ni and Au+Au collisions at 1.5 $A\cdot$GeV.
{\it Preliminary results!}.
}
\label{Massaction}
\end{figure}

This relation also explains the measured constant ratio of $K^-/K^+$ 
with centrality (Fig.~\ref{KPKM_SIS_RHIC}) as the pion multiplicity does not 
vary with centrality.
Figure~\ref{Massaction} demonstrates the constancy of the $K^-/K^+$
ratio and of the pion multiplicity with $A_{part}$ 
for Ni+Ni and Au+Au collisions at 1.5 $A\cdot$GeV~\cite{Marc,AF,FU}.
It turns out that these ratios do not even
depend on the choice of the collision system. 
The right part of this figure exhibits the double ratio
$([K^-]/[K^+])/([M(\pi^0 + \pi^-)]/A_{part})$ which shows only a minor 
deviation from a horizontal line. This result can be taken as an argument that
this specific channel might be not far from chemical equilibrium.

\begin{figure}
\begin{minipage}[t]{5.3cm}
\mbox{\epsfig{width=9.6cm,file=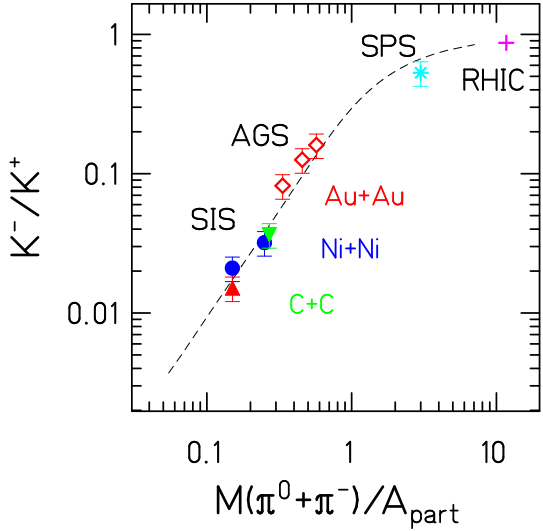}}
\end{minipage}
\hfill
\begin{minipage}[b]{7.8cm}
\vspace*{-5cm}
\caption{The $K^-/K^+$ ratio as a function of the
pion multiplicity $M(\pi^-+\pi^0)/A_{part}$ as a test of the law of mass action.
{\it Preliminary data.} The dashed line shows the prediction of the statistical 
model.}
\label{Massaction_SIS_RHIC}
\end{minipage}
\end{figure}

Next we test in Fig.~\ref{Massaction_SIS_RHIC} 
the validity of the law of mass action by plotting the
$K^-/K^+$ ratio as a function of the pion multiplicity 
$M(\pi^0 + \pi^-)/A_{part}$ at incident energies from SIS up to RHIC.
At SIS and AGS energies the direct relation holds, i.e.~the $K^-/K^+$ ratio
rises with $M(\pi^0 + \pi^-)/A_{part}$. 
At SPS and RHIC
energies $K^-$ are obviously produced by other channels, 
i.e.~$K^+ K^-$ pair production. This change of the dominating channel is well
reproduced by the statistical model (dashed line in 
Fig.~\ref{Massaction_SIS_RHIC}).

\subsection{Maximum relative strangeness content
 in heavy ion collisions around 30 A$\cdot$GeV}

\begin{figure}
\begin{minipage}[t]{5.3cm}
\mbox{\epsfig{width=7.6cm,file=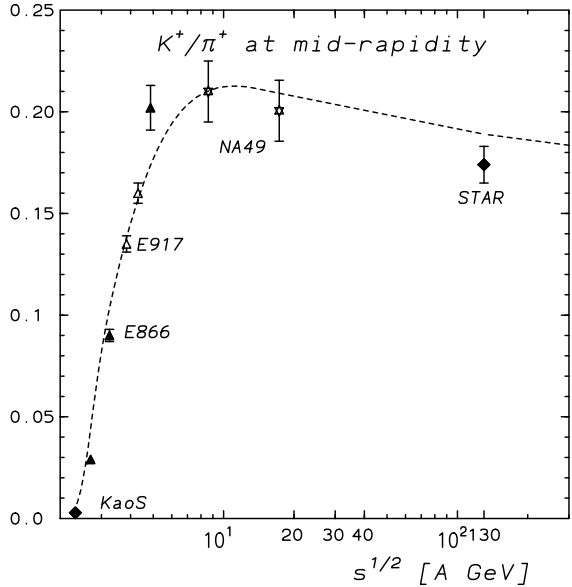}}
\end{minipage}
\hfill
\begin{minipage}[b]{7.8cm}
\vspace*{-3.6cm}
\caption{$K^+/\pi^+$ ratio obtained around midrapidity as a
function of $\sqrt s$ from the various experiments. 
The dashed
line shows the results of the statistical model in complete equilibrium.}
\label{KP_PI_ratio}
\end{minipage}
\end{figure}

The experimental data from heavy ion collisions show that
the $K^+/\pi^+$ ratio rises from SIS up to AGS but it is larger
for AGS than at the highest CERN-SPS energies
\cite{CLEY98,E802,dunlop,blume,bearden} 
and even at RHIC \cite{Star} as shown in Fig.~\ref{KP_PI_ratio}.

 This behavior is of
particular interest as it could signal the appearance of new
dynamics for strangeness production in high energy collisions. It
was  even conjectured~\cite{gazdzicki}
 that this property could
indicate   an energy    threshold  for  quark-gluon plasma
formation in relativistic heavy ion collisions.

In the following we
analyze the energy dependence of strange to non-strange
particle  ratios in the framework of a hadronic
statistical model. In the whole  energy range, the
hadronic yields  observed in heavy ion collisions resemble those
of a population in chemical equilibrium along a unified freeze-out
curve determined by the condition of fixed energy/particle
$\simeq$ 1 GeV \cite{CLEY98} providing a relation between 
the temperature $T$ and the baryon chemical potential $\mu_B$.
As the beam energy increases $T$ rises and $\mu_B$ is slightly reduced.
Above AGS energies, $T$ exhibits only a
moderate change and converges to
its maximal value in the range of 160 to 180 MeV, while $\mu_B$ is strongly 
decreasing.

Instead of studying the $K^+/\pi^+$ ratio we use the ratios of
 strange to non-strange particle
multiplicities (Wroblewski
factor)~\cite{wroblewski}
defined as
$$
\lambda_s \equiv {2\bigl<s\bar{s}\bigr>\over
\bigl<u\bar{u}\bigr> + \bigl<d\bar{d}\bigr>}
$$
where the  quantities in angular brackets refer to the number of
newly formed quark-antiquark pairs, i.e.~it excludes all
quarks that were present in the target and the projectile.

\begin{figure}
\begin{minipage}[t]{5.0cm}
\mbox{\epsfig{width=8.8cm,file=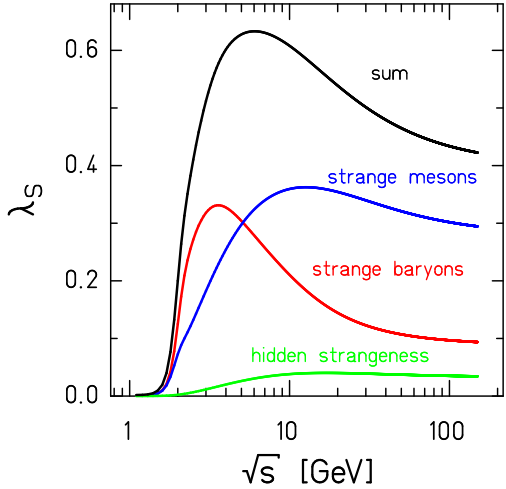}}
\end{minipage}
\hfill
\begin{minipage}[b]{7.8cm}
\vspace*{-4.9cm}
\caption{Contributions to the Wroblewski factor $\lambda_s$ (for definition
see text) from strange baryons, 
strange mesons, and mesons with hidden strangeness. 
The sum of all contributions is given by the full line.}
\label{Wrob_composition}
\end{minipage}
\end{figure}

Applying the statistical model to particle production in heavy ion
collisions calls for the  use of the canonical ensemble
to treat the number of strange particles
particularly  for data in the energy range
from SIS up to AGS \cite{CLE99,ko} as mentioned before.
The calculations for Au-Au and
Pb-Pb collisions  are performed using
a canonical correlation volume defined above and 
given by a radius of $\sim 7$ fm
determined in \cite{CLE99}.
The quark content used in the Wroblewski factor is determined at the moment
of {\it {chemical freeze-out}}, i.e.~from the hadrons and especially, hadronic
resonances, before they decay.
This ratio is thus not an easily measurable  observable
unless one can reconstruct all resonances from the final-state
particles.  The results are shown in Fig.~\ref{Wrob_composition} 
as a function of $\sqrt{s}$.

The solid line (marked ``sum'') in Fig.~\ref{Wrob_composition}
describes the statistical-model
calculations in complete equilibrium along the unified freeze-out
curve~\cite{CLEY98} and with the energy-dependent parameters $T$ and $\mu_B$.
From Fig.~\ref{Wrob_composition} we conclude that around 30 $A\cdot$GeV
laboratory energy the relative strangeness content in heavy ion
collisions reaches a
 clear and well pronounced maximum.
The Wroblewski factor  decreases towards higher incident energies
and reaches a limiting value of about 0.43.
For details see Ref.~\cite{max_strange}.

\begin{figure}
\begin{minipage}[t]{5.3cm}
\mbox{\epsfig{width=9.0cm,file=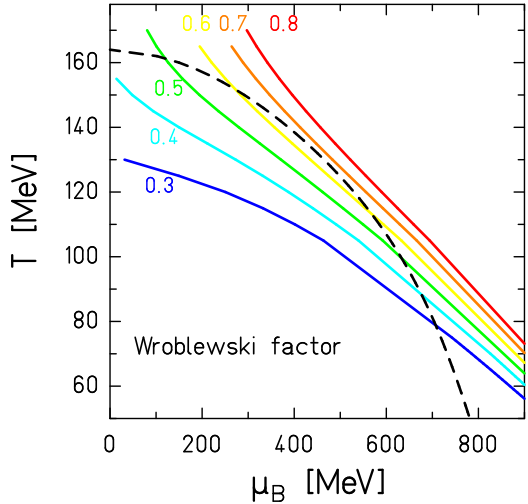}}
\end{minipage}
\hfill
\begin{minipage}[b]{7.8cm}
\vspace*{-4.0cm}
\caption{Lines of constant Wroblewski factor $\lambda_s$ (for
definition  see text) in the $T-\mu_B$ plane (solid lines)
together with the freeze-out curve (dashed line)~\protect\cite{CLEY98}.}
\label{Wrob_T_MUB}
\end{minipage}
\end{figure}

The appearance of the maximum can be traced  to the specific
dependence of $\mu_B$ and $T$ on the beam energy.
Figure~\ref{Wrob_T_MUB} shows
values of constant $\lambda_s$  in the $T-\mu_B$
plane. As expected $\lambda_s$ rises with increasing $T$ for fixed
$\mu_B$.
Following the chemical freeze-out curve, shown as a full dashed line in
Fig.~\ref{Wrob_T_MUB}, one can see that
 $\lambda_s$ rises quickly from SIS to AGS energies,
then reaches  a maximum around $\mu_B\approx 500$ MeV
and $T\approx 130$ MeV.
These freeze-out parameters correspond to
30 GeV laboratory energy. At higher incident
energies the increase in $T$ becomes negligible but $\mu_B$ keeps
on decreasing and as a consequence $\lambda_s$ also decreases.

The importance of finite baryon density on the
behavior of $\lambda_s$ is demonstrated in  Fig.~\ref{Wrob_composition} showing
separately the  contributions to $\left<s\bar{s}\right>$
coming from strange baryons, from strange mesons and from hidden strangeness, 
i.e.~from hadrons  like $\phi$ and $\eta$.
As can be seen in Fig.~\ref{Wrob_composition},
the origin of the maximum in the Wroblewski ratio can be traced  to the 
contribution of strange baryons.
This channel dominates at low $\sqrt{s}$ and loses
importance at  high incident energies.
Even strange mesons exhibit a broad maximum. This is due to the
presence of associated production of e.g.~kaons together with
hyperons. 

 The energy dependence of the
$K^+/\pi^+$ ratio measured at midrapidity
is shown in Fig.~\ref{KP_PI_ratio}.
The model  gives an excellent description of the data, showing
a broad maximum at the same energy as the one
seen in the Wroblewski factor.
In general, of course, statistical-model calculations
 should be compared with
4$\pi$-integrated results since strangeness does not have to be
conserved in a limited portion of phase space.
A drop in this
ratio for 4$\pi$ yields has been reported from preliminary results
of the NA49 collaboration  at 158 AGeV~\cite{blume}. This decrease
is, however, not reproduced by the statistical model
 without further modifications, e.g.~by introducing an additional
parameter $\gamma_s\sim 0.7$ \cite{becattini}.
This point might be clearer when data at other beam
energies  will become available.

\section{Summary}

Strange particle production in heavy ion collisions close to threshold 
can be described by a statistical model in canonical formulation, i.e.~using
local strangeness conservation. This approach is able to explain
several features of $K^+$ and $K^-$ production at SIS energies.

While for $K^+$ production it remains open whether and how chemical
equlibrium can be reached, the situation for $K^-$ is quite different.
It is shown that the strangeness exchange process 
$\pi \Lambda \rightleftharpoons N + K^-$ is the dominant channel for 
$K^-$ production at SIS and likely also at AGS energies. This is demonstrated
by applying the corresponding law of mass action. Theoretical studies
confirm this interpretation. 

Using the energy dependence of the parameters $T$ and $\mu_B$
we have  shown
that the statistical-model description of relativistic heavy ion
collisions predicts that the yields of strange to nonstrange
particles reaches  a well defined maximum near 30 GeV lab energy.
It is demonstrated that this maximum is due to the specific shape
of the freeze-out curve in the $T-\mu_B$ plane. In particular,  a
very steep decrease of the baryon chemical potential with
increasing energy causes a corresponding decline of relative
strangeness content in systems created in heavy ion
collisions above lab energies of 30 GeV. The saturation in $T$,
necessary for this result, might be connected to the fact that
hadronic temperatures cannot exceed the critical temperature
$T_c\simeq$ 170 MeV for the phase transition to the QGP as found
in solutions of QCD on the lattice.

\end{document}